\pdfoutput=1

\documentclass[12pt,a4paper]{article}

\usepackage{ifthen} 
\newboolean{pdflatex}
\setboolean{pdflatex}{true} 

\newboolean{articletitles}
\setboolean{articletitles}{true} 

\newboolean{uprightparticles}
\setboolean{uprightparticles}{false} 

\newboolean{inbibliography}
\setboolean{inbibliography}{false} 

\usepackage[top=1in, bottom=1.25in, left=1in, right=1in]{geometry}

\usepackage{microtype}
\usepackage{lineno}  
\usepackage{xspace} 
\usepackage{caption} 

\usepackage{graphicx}  
\usepackage{color}
\usepackage{colortbl}
\graphicspath{{./figs/}} 

\usepackage{amsmath} 
\usepackage{amssymb}
\usepackage{amsfonts}
\usepackage{upgreek} 

\newcommand*\patchAmsMathEnvironmentForLineno[1]{%
\expandafter\let\csname old#1\expandafter\endcsname\csname #1\endcsname
\expandafter\let\csname oldend#1\expandafter\endcsname\csname
end#1\endcsname
 \renewenvironment{#1}%
   {\linenomath\csname old#1\endcsname}%
   {\csname oldend#1\endcsname\endlinenomath}%
}
\newcommand*\patchBothAmsMathEnvironmentsForLineno[1]{%
  \patchAmsMathEnvironmentForLineno{#1}%
  \patchAmsMathEnvironmentForLineno{#1*}%
}
\AtBeginDocument{%
\patchBothAmsMathEnvironmentsForLineno{equation}%
\patchBothAmsMathEnvironmentsForLineno{align}%
\patchBothAmsMathEnvironmentsForLineno{flalign}%
\patchBothAmsMathEnvironmentsForLineno{alignat}%
\patchBothAmsMathEnvironmentsForLineno{gather}%
\patchBothAmsMathEnvironmentsForLineno{multline}%
\patchBothAmsMathEnvironmentsForLineno{eqnarray}%
}

\usepackage{hyperref}    
\usepackage[all]{hypcap} 

\usepackage{xspace} 
\usepackage{upgreek}

\def\lhcb {\mbox{LHCb}\xspace}

\def\MagUp {\mbox{\em Mag\kern -0.05em Up}\xspace}

\ifthenelse{\boolean{uprightparticles}}%
{

 \def\Peta        {\ensuremath{\upeta}\xspace}

 \def\Pmu         {\ensuremath{\upmu}\xspace}

 \def\Ppsi        {\ensuremath{\uppsi}\xspace}

 \def\PDelta      {\ensuremath{\Delta}\xspace}                 
 \def\PXi      {\ensuremath{\Xi}\xspace}                 
 \def\PLambda      {\ensuremath{\Lambda}\xspace}                 
 \def\PSigma      {\ensuremath{\Sigma}\xspace}                 
 \def\POmega      {\ensuremath{\Omega}\xspace}                 
 \def\PUpsilon      {\ensuremath{\Upsilon}\xspace}                 
 
 \def\PB      {\ensuremath{\mathrm{B}}\xspace}                 
                  
 \def\PD      {\ensuremath{\mathrm{D}}\xspace}

 \def\PJ      {\ensuremath{\mathrm{J}}\xspace}                 
 \def\PK      {\ensuremath{\mathrm{K}}\xspace}

 \def\Pb      {\ensuremath{\mathrm{b}}\xspace}                 
 \def\Pc      {\ensuremath{\mathrm{c}}\xspace}

 \def\Pi      {\ensuremath{\mathrm{i}}\xspace}

}
{

 \def\Peta        {\ensuremath{\eta}\xspace}

 \def\Pmu         {\ensuremath{\mu}\xspace}

 \def\Ppsi        {\ensuremath{\psi}\xspace}                 
                  
 \mathchardef\PDelta="7101
 \mathchardef\PXi="7104
 \mathchardef\PLambda="7103
 \mathchardef\PSigma="7106
 \mathchardef\POmega="710A
 \mathchardef\PUpsilon="7107
                  
 \def\PB      {\ensuremath{B}\xspace}                 
                  
 \def\PD      {\ensuremath{D}\xspace}

 \def\PJ      {\ensuremath{J}\xspace}                 
 \def\PK      {\ensuremath{K}\xspace}

 \def\Pb      {\ensuremath{b}\xspace}                 
 \def\Pc      {\ensuremath{c}\xspace}

 \def\Pi      {\ensuremath{i}\xspace}

}

\makeatletter
\ifcase \@ptsize \relax
  \newcommand{\miniscule}{\@setfontsize\miniscule{4}{5}}
\or
  \newcommand{\miniscule}{\@setfontsize\miniscule{5}{6}}
\or
  \newcommand{\miniscule}{\@setfontsize\miniscule{5}{6}}
\fi
\makeatother

\DeclareRobustCommand{\optbar}[1]{\shortstack{{\miniscule (\rule[.5ex]{1.25em}{.18mm})}
  \\ [-.7ex] $#1$}}

\def\mup        {{\ensuremath{\Pmu^+}}\xspace}
\def\mun        {{\ensuremath{\Pmu^-}}\xspace} 

\def\cquark    {{\ensuremath{\Pc}}\xspace}

\def\bquark    {{\ensuremath{\Pb}}\xspace}

  \def\Kbar    {{\kern 0.2em\overline{\kern -0.2em \PK}{}}\xspace}

\def\KorKbar    {\kern 0.18em\optbar{\kern -0.18em K}{}\xspace}

  \def\Dbar    {{\kern 0.2em\overline{\kern -0.2em \PD}{}}\xspace}

\def\DorDbar    {\kern 0.18em\optbar{\kern -0.18em D}{}\xspace}

\def\B       {{\ensuremath{\PB}}\xspace}
\def\Bbar    {{\ensuremath{\kern 0.18em\overline{\kern -0.18em \PB}{}}}\xspace}

\def\BorBbar    {\kern 0.18em\optbar{\kern -0.18em B}{}\xspace}

\def\Bu      {{\ensuremath{\B^+}}\xspace}

\def\Bp      {{\ensuremath{\Bu}}\xspace}

\def\jpsi     {{\ensuremath{{\PJ\mskip -3mu/\mskip -2mu\Ppsi\mskip 2mu}}}\xspace}

\def\etac     {{\ensuremath{\Peta_\cquark}}\xspace}

  \def\Y#1S{\ensuremath{\PUpsilon{(#1S)}}\xspace}

\def\Lz          {{\ensuremath{\PLambda}}\xspace}
\def\Lbar        {{\ensuremath{\kern 0.1em\overline{\kern -0.1em\PLambda}}}\xspace}
\def\LorLbar    {\kern 0.18em\optbar{\kern -0.18em \PLambda}{}\xspace}

\def\to                 {\ensuremath{\rightarrow}\xspace}

\def\AT#1     {\ensuremath{A_{\mathrm{T}}^{#1}}\xspace}           

\def\C#1      {\ensuremath{\mathcal{C}_{#1}}\xspace}                       
\def\Cp#1     {\ensuremath{\mathcal{C}_{#1}^{'}}\xspace}                    
\def\Ceff#1   {\ensuremath{\mathcal{C}_{#1}^{\mathrm{(eff)}}}\xspace}        
\def\Cpeff#1  {\ensuremath{\mathcal{C}_{#1}^{'\mathrm{(eff)}}}\xspace}       
\def\Ope#1    {\ensuremath{\mathcal{O}_{#1}}\xspace}                       
\def\Opep#1   {\ensuremath{\mathcal{O}_{#1}^{'}}\xspace}                    

\newcommand{\tev}{\ifthenelse{\boolean{inbibliography}}{\ensuremath{~T\kern -0.05em eV}\xspace}{\ensuremath{\mathrm{\,Te\kern -0.1em V}}}\xspace}
\newcommand{\gev}{\ensuremath{\mathrm{\,Ge\kern -0.1em V}}\xspace}
\newcommand{\mev}{\ensuremath{\mathrm{\,Me\kern -0.1em V}}\xspace}
\newcommand{\kev}{\ensuremath{\mathrm{\,ke\kern -0.1em V}}\xspace}
\newcommand{\ev}{\ensuremath{\mathrm{\,e\kern -0.1em V}}\xspace}
\newcommand{\gevc}{\ensuremath{{\mathrm{\,Ge\kern -0.1em V\!/}c}}\xspace}
\newcommand{\mevc}{\ensuremath{{\mathrm{\,Me\kern -0.1em V\!/}c}}\xspace}
\newcommand{\gevcc}{\ensuremath{{\mathrm{\,Ge\kern -0.1em V\!/}c^2}}\xspace}
\newcommand{\gevgevcccc}{\ensuremath{{\mathrm{\,Ge\kern -0.1em V^2\!/}c^4}}\xspace}
\newcommand{\mevcc}{\ensuremath{{\mathrm{\,Me\kern -0.1em V\!/}c^2}}\xspace}

\def\mum  {\ensuremath{{\,\upmu\mathrm{m}}}\xspace}

\def\gsim{{~\raise.15em\hbox{$>$}\kern-.85em
          \lower.35em\hbox{$\sim$}~}\xspace}
\def\lsim{{~\raise.15em\hbox{$<$}\kern-.85em
          \lower.35em\hbox{$\sim$}~}\xspace}

\def\sPlot{\mbox{\em sPlot}\xspace}

\def\ptot       {\mbox{$p$}\xspace}
\def\pt         {\mbox{$p_{\mathrm{ T}}$}\xspace}

\def\evtgen     {\mbox{\textsc{EvtGen}}\xspace}

\def\geant      {\mbox{\textsc{Geant4}}\xspace}

\def\photos     {\mbox{\textsc{Photos}}\xspace}

\def\pythia     {\mbox{\textsc{Pythia}}\xspace}

\def\tell1  {TELL1\xspace}
\def\ukl1   {UKL1\xspace}

\usepackage{cite} 
\usepackage{mciteplus}

\usepackage{longtable} 

\begin{document}

\renewcommand{\thefootnote}{\fnsymbol{footnote}}
\setcounter{footnote}{1}


\begin{titlepage}
\pagenumbering{roman}

\vspace*{-1.5cm}
\centerline{\large EUROPEAN ORGANIZATION FOR NUCLEAR RESEARCH (CERN)}
\vspace*{1.5cm}
\noindent
\begin{tabular*}{\linewidth}{lc@{\extracolsep{\fill}}r@{\extracolsep{0pt}}}
\ifthenelse{\boolean{pdflatex}}
{\vspace*{-2.7cm}\mbox{\!\!\!\includegraphics[width=.14\textwidth]{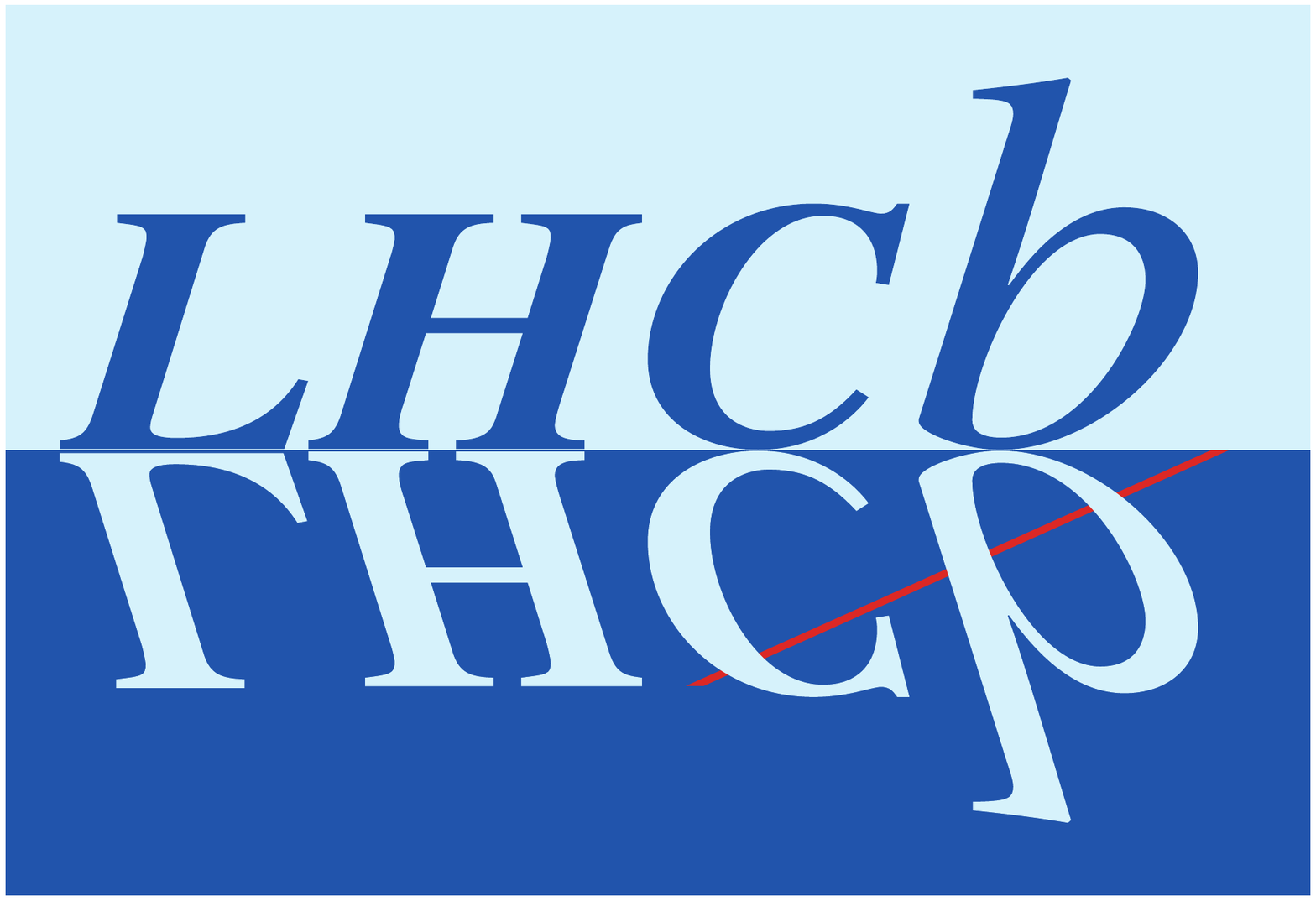}} & &}%
{\vspace*{-1.2cm}\mbox{\!\!\!\includegraphics[width=.12\textwidth]{lhcb-logo.eps}} & &}%
\\
 & & CERN-EP-2016-175 \\  
 & & LHCb-PAPER-2016-016 \\  
 & & 21 July 2016 \\ 
 & & \\
\end{tabular*}

\vspace*{1.0cm}

{\normalfont\bfseries\boldmath\huge
\begin{center}
  Observation of $\eta_{c}(2S) \to p \bar p$ and search for
  $X(3872) \to
p \bar p$ decays
\end{center}
}

\vspace*{0.25cm}

\begin{center}
The LHCb collaboration\footnote{Authors are listed at the end of this paper.}
\end{center}

\vspace*{0.25cm}

\begin{abstract}
  \noindent
The first observation of the decay $\eta_{c}(2S) \to p \bar p$ is
reported using proton-proton collision data corresponding to an integrated luminosity of $3.0\rm \, fb^{-1}$
recorded by the LHCb experiment at centre-of-mass energies of 7 and 8 TeV. The $\eta_{c}(2S)$ resonance is
produced in the decay $B^{+} \to [c\bar c] K^{+}$. The product of 
branching fractions normalised to
that for the $\jpsi$ intermediate state, ${\cal R}_{\eta_{c}(2S)}$, is measured to be
\begin{align*}
{\cal R}_{\eta_{c}(2S)}\equiv\frac{{\mathcal B}(B^{+} \to \eta_{c}(2S) K^{+}) \times {\mathcal B}(\eta_{c}(2S) \to
p \bar p)}{{\mathcal B}(B^{+} \to \jpsi K^{+}) \times {\mathcal B}(\jpsi \to
p \bar p)}  =~& (1.58 \pm 0.33 \pm 0.09)\times 10^{-2},
\end{align*}
where the first uncertainty is statistical and the second systematic. 
No signals for the decays $B^{+} \to X(3872) (\to p \bar p)
K^{+}$ and  $B^{+} \to \psi(3770) (\to p \bar p)
K^{+}$ are seen, and the 95\% confidence level upper limits on their
relative branching ratios are
found to be ${\cal R}_{X(3872)}<0.25\times10^{-2}$ and  ${\cal R}_{\psi(3770))}<0.10$.
In addition, the mass differences between the $\eta_{c}(1S)$ and the
$\jpsi$ states, between the $\eta_{c}(2S)$ and the $\psi(2S)$
states, and the natural width of the $\eta_{c}(1S)$ are measured as
\begin{align*}
M_{\jpsi}  - M_{\eta_{c}(1S)} =~&  110.2 \pm 0.5 \pm
0.9  \mev,\\
M_{\psi(2S)}  -M_{\eta_{c}(2S)} =~ & 52.5 \pm 1.7  \pm 0.6 \mev,\\
\Gamma_{\eta_{c}(1S)} =~& 34.0 \pm 1.9  \pm 1.3 \mev.
\end{align*}
\end{abstract}

\vspace*{1.0cm}

\begin{center}
Submitted to Phys.~Lett.~B 
\end{center}

\vspace{\fill}

{\footnotesize 
\centerline{\copyright~CERN on behalf of the \lhcb collaboration, licence \href{http://creativecommons.org/licenses/by/4.0/}{CC-BY-4.0}.}}
\vspace*{2mm}

\end{titlepage}


\newpage
\setcounter{page}{2}
\mbox{~}

\cleardoublepage


\renewcommand{\thefootnote}{\arabic{footnote}}
\setcounter{footnote}{0}



\pagestyle{plain} 
\setcounter{page}{1}
\pagenumbering{arabic}

\section{Introduction}
\label{sec:Introduction}
Charmonium has proved to be a remarkable laboratory for the study of quantum
chromodynamics in the non-perturbative regime. By comparing theoretical
predictions with experimental results one can verify and tune the parameters of theoretical models in order to improve the accuracy of the predictions. In addition, in recent
years, many exotic charmonium-like states have been observed, renewing
interest in charmonium spectroscopy above the open-charm
threshold~\cite{Brambilla:2010cs,Eidelman:2014}. 
The $B^{+} \to p \bar p K^{+}$ decay\footnote{The inclusion of
  charge-conjugate modes is implied throughout the paper.} offers a clean environment to study
intermediate resonances, such as charmonium and charmonium-like states
decaying to $p\bar p$.
The presence of $p\bar p$ in the final state allows intermediate
states of any quantum number to be studied.

The first radial excitation $\eta_{c}(2S)$ of the charmonium ground
state $\eta_{c}(1S)$ was observed at the \B factories~\cite{Choi:2002na, 
Aubert:2003pt, Asner:2003wv} and, to date,
only a few of its decay modes have been observed.
The BESIII collaboration has recently searched for the $\eta_{c}(2S)
\to p\bar p$ decay in $\psi(2S)$ radiative
transitions~\cite{Ablikim:2013hdv}, and set an upper limit on the
product of branching fractions ${\mathcal B}(\psi(3686) \to 
\eta_{c}(2S)\gamma) \times {\mathcal B}(\eta_{c}(2S) \to p \bar p)$.

The $\eta_{c}(1S)$ state is the lowest-lying S-wave spin-singlet charmonium state
and has been observed in various processes.
The measurements of  the $\eta_{c}(1S)$ mass and width in radiative charmonium
transitions show a tension with those determined in different processes such as photon-photon fusion and \B decays~\cite{PDG2014}. Detailed investigations of the
line shape of the magnetic dipole transition by the
KEDR~\cite{Anashin:2014wva} and CLEO~\cite{Mitchell:2008aa}
collaborations indicate that additional factors modify the na\"\i ve
$k^3$ dependence on the photon momentum, $k$,  assumed in earlier measurements. This
would affect the measurements of the mass and width in radiative
charmonium transitions.

In this paper, the first observation of $\eta_{c}(2S) \to p \bar
p$ decay and a search for $\psi(3770) \to p \bar p$ and $X(3872) \to p \bar p$ decays are
reported. The measurements of the branching fractions are relative to that of the $B^{+}
\to \jpsi (\to p \bar p) K^{+} $ decay.
Additional measurements of the $\eta_c(1S)$ and $\eta_c(2S)$ mass and the $\eta_c(1S)$ width are reported.
This new measurement of the $\eta_c(1S)$ resonance parameters in
exclusive $B^{+} \to [c
\bar c] K^{+}$ decays, where $[c\bar c]$ stands for a generic
charmonium resonance, is independent of the above-mentioned
line-shape complications.

\section{Detector and simulation}
\label{sec:Detector}
The \lhcb detector~\cite{Alves:2008zz,LHCb-DP-2014-002} is a single-arm forward
spectrometer covering the \mbox{pseudorapidity} range $2<\eta <5$,
designed for the study of particles containing \bquark or \cquark
quarks. The detector includes a high-precision tracking system
consisting of a silicon-strip vertex detector surrounding the $pp$
interaction region, a large-area silicon-strip detector located
upstream of a dipole magnet with a bending power of about
$4{\mathrm{\,Tm}}$, and three stations of silicon-strip detectors and straw
drift tubes placed downstream of the magnet.
The tracking system provides a measurement of momentum, \ptot, of charged particles with
a relative uncertainty that varies from 0.5\% at low momentum~\footnote{Natural units with $c=1$ are used throughout the
paper} to 1.0\% at 200\gev.
The minimum distance of a track to a primary vertex~(PV), the impact
parameter (IP), is measured with a resolution of $(15+29/\pt)\mum$,
where \pt is the component of the momentum transverse to the beam, in\,\gev.
Different types of charged hadrons are distinguished using information
from two ring-imaging Cherenkov
detectors. 
The online event selection is performed by a trigger, 
which consists of a hardware stage, based on information from the calorimeter and muon
systems, followed by a software stage, which applies full event
reconstruction.

At the hardware trigger stage, events are required to have high
transverse energy in the calorimeters. For hadrons,
the transverse energy threshold is 3.5\gev.
The software trigger requires the presence of a two-, three- or four-track
secondary vertex with significant displacement from the primary
$pp$ interaction vertices. At least one charged particle
must have $\pt$ larger than $1.7\gev$ and be
inconsistent with originating from a PV.
A multivariate algorithm~\cite{BBDT} is used for
the identification of secondary vertices consistent with the decay
of a \bquark hadron.

Simulated decays of $B^+ \to \eta_{c}(2S) (\to p \bar p) K^{+}$, $B^+ \to X(3872) (\to p \bar p)
K^{+}$, $B^+ \to \psi(2S) (\to p \bar p)
K^{+}$  and $B^{+} \to p \bar p K^{+}$, generated uniformly in
phase space, are used to
optimise the signal selection and to evaluate the ratio of the efficiencies for each considered
channel with respect to the $B^+ \to
\jpsi (\to p \bar p) K^{+}$  mode. In the simulation, $pp$
collisions are generated using \pythia\!8~\cite{Sjostrand:2007gs, *Sjostrand:2006za}  with a specific \lhcb
configuration~\cite{LHCb-PROC-2010-056}.  Decays of hadronic particles
are described by \evtgen~\cite{Lange:2001uf}, in which final-state
radiation is simulated using \photos~\cite{Golonka:2005pn}. The
interaction of the generated particles with the detector, and its
response, are implemented using the \geant
toolkit~\cite{Allison:2006ve, *Agostinelli:2002hh} as described in Ref.~\cite{LHCb-PROC-2011-006}.
\section{Event selection}
\label{sec:eventselection}
The selection of the $B^{+}$ candidates is done in two stages. A loose preselection is based on track quality,
momentum, transverse momentum, impact parameter of the $B^+$ candidate
and its daughters, $B^+$
flight distance, and particle identification (PID) of the $p$ and $\bar p$ candidates. 
The reconstructed $B^+$ candidates are required to have a $p\bar p K^{+}$
invariant mass in the range $5.08 \-- 5.68 \gev$. The asymmetric invariant mass range
around the known $B^+$ mass is chosen to select $B^+ \to p
\bar p \pi^{+}$ candidates also. 

The reconstructed candidates that meet the above criteria are further filtered using a boosted
decision tree (BDT) algorithm~\cite{Breiman,Roe}. The BDT is trained
on a signal sample of simulated $B^{+} \to p \bar p K^{+}$ decays and a background sample of data taken from the
upper $B^{+}$-mass sideband in the range $5.34 \-- 5.48 \gev$. The upper sideband is exploited to avoid partially reconstructed
background. 
Input quantities include
variables related to the $B^{+}$ candidate
and its daughter particles, \Bp decay vertex quality and
its displacement from the PV, \Bp flight direction inferred from the
two vertex positions, \Bp momentum and final-state particle identification. 
The selection criterion on
the BDT response is chosen by maximising the significance of the $\chi_{c1}\to p \bar{p}$ signal yield in data. The number of
events from this well-known transition
provides a control sample comparable in size to that of the $\eta_{c}(2S)$. 

\section{Invariant mass spectra and event yields}
\label{sec:eventyields}
An extended unbinned maximum likelihood fit is performed to the
$p\bar p K^{+}$ invariant mass distribution.
The backgrounds observed in the $p\bar p K^{+}$ mass distribution are subtracted using
the \sPlot technique~\cite{Pivk:2004ty} to extract the $p\bar p$ mass spectrum in $B^{+}\to p\bar p K^{+}$ decays.
Signal yields for the resonant contributions are then determined from an
extended unbinned maximum likelihood fit to the $p\bar p$ mass
spectrum. To improve the
$p\bar p$ invariant mass resolution, the fit to the $B^{+}$ decay
vertex is performed with the $B^{+}$
mass constrained to the known value~\cite{PDG2014} and the $B^{+}$ candidate
pointing to the PV~\cite{Hulsbergen:2005pu}. The $p\bar p$ mass spectrum is also used to
determine the mass differences $M_{\jpsi}-M_{\eta_{c}(1S)}$ and
$M_{\psi(2S)}-M_{\eta_{c}(2S)}$ and the natural width of the
$\eta_{c}(1S)$ state. In order to have accurate mass measurements, a calibration is applied to the
momenta of the final-state particles. Large samples of $B^{+} \to
\jpsi K^{+}$ decays with $\jpsi \to \mup\mun$ are used to calibrate the momentum scale of the spectrometer~\cite{LHCb-PAPER-2012-048}.

\begin{figure}[b!!!]
\centering
\includegraphics[scale=0.65]{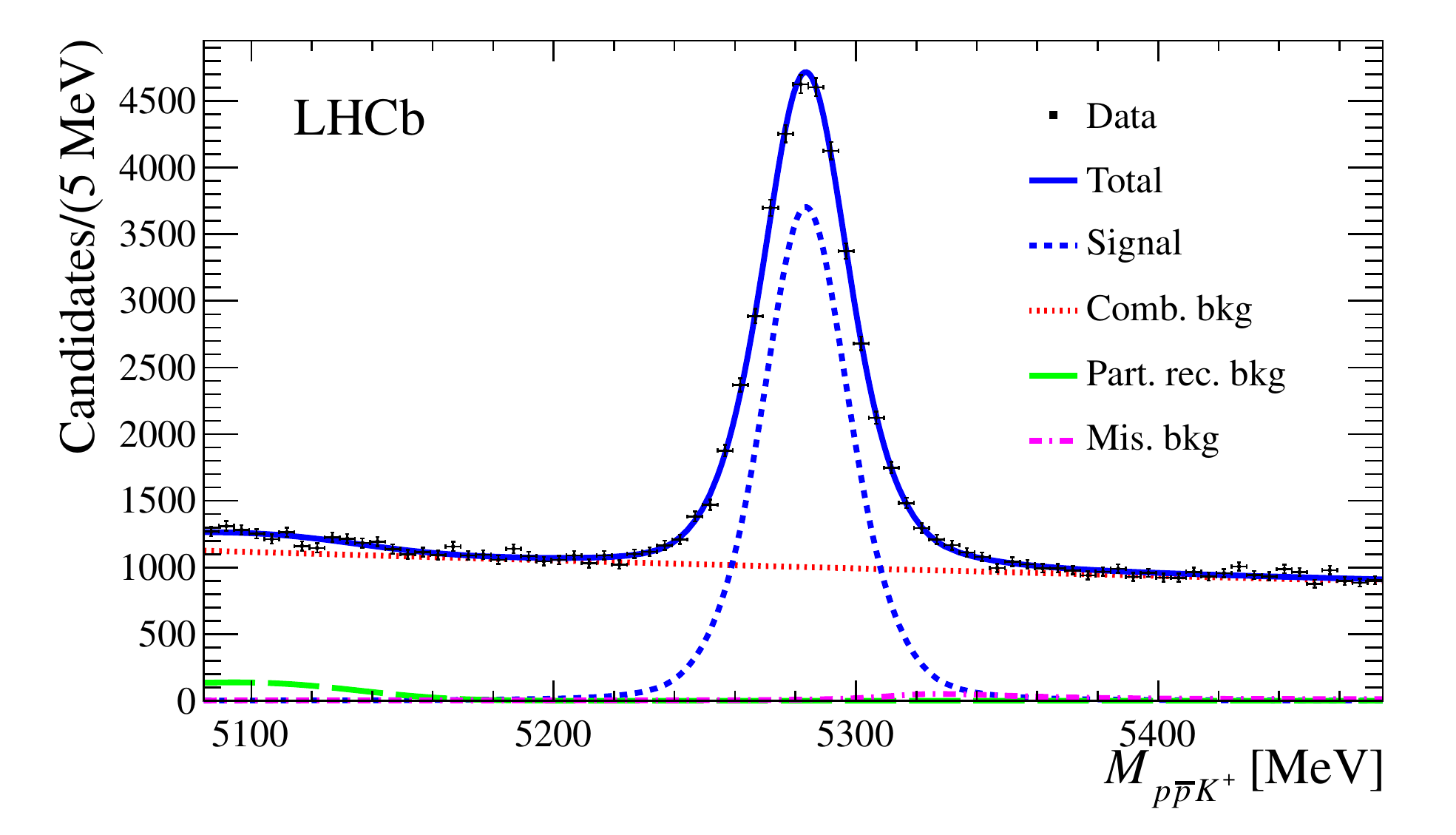}
\caption{Invariant mass spectrum of the $p \bar{p} K^+$
  candidates. The total fit curve and individual fit components are
  superimposed on the data.}
\label{fig:massfit}
\end{figure}

The $p \bar p K^{+}$ invariant mass distribution is shown in
Fig.~\ref{fig:massfit}. The signal peak is parameterised using an Apollonios
probability density function (PDF)~\cite{Santos:2013gra}. The yield,
mean and resolution are allowed to
vary freely in the fit, while the tail parameters are fixed to the values obtained from simulation.
The combinatorial background component is parameterised by an
exponential function. Partially reconstructed background due to $B^{+}
\to p \bar p K^{+} \pi^{0}$ decays
is parameterised using an ARGUS PDF~\cite{Albrecht:1994tb} convolved
with a Gaussian resolution function with parameters fixed to the values obtained from simulation.
The misidentified background due to $B^{+} \to p \bar p \pi^{+}$ decays,
where the charged pion is misidentified as a kaon, is parameterised
with a bifurcated Gaussian PDF~\cite{delAmoSanchez:2010ae} and parameters fixed to the values obtained from simulation. The yields of partially reconstructed and misidentified backgrounds are determined from data.

\begin{figure}[b!!!]
\centering
\includegraphics[scale=0.8]{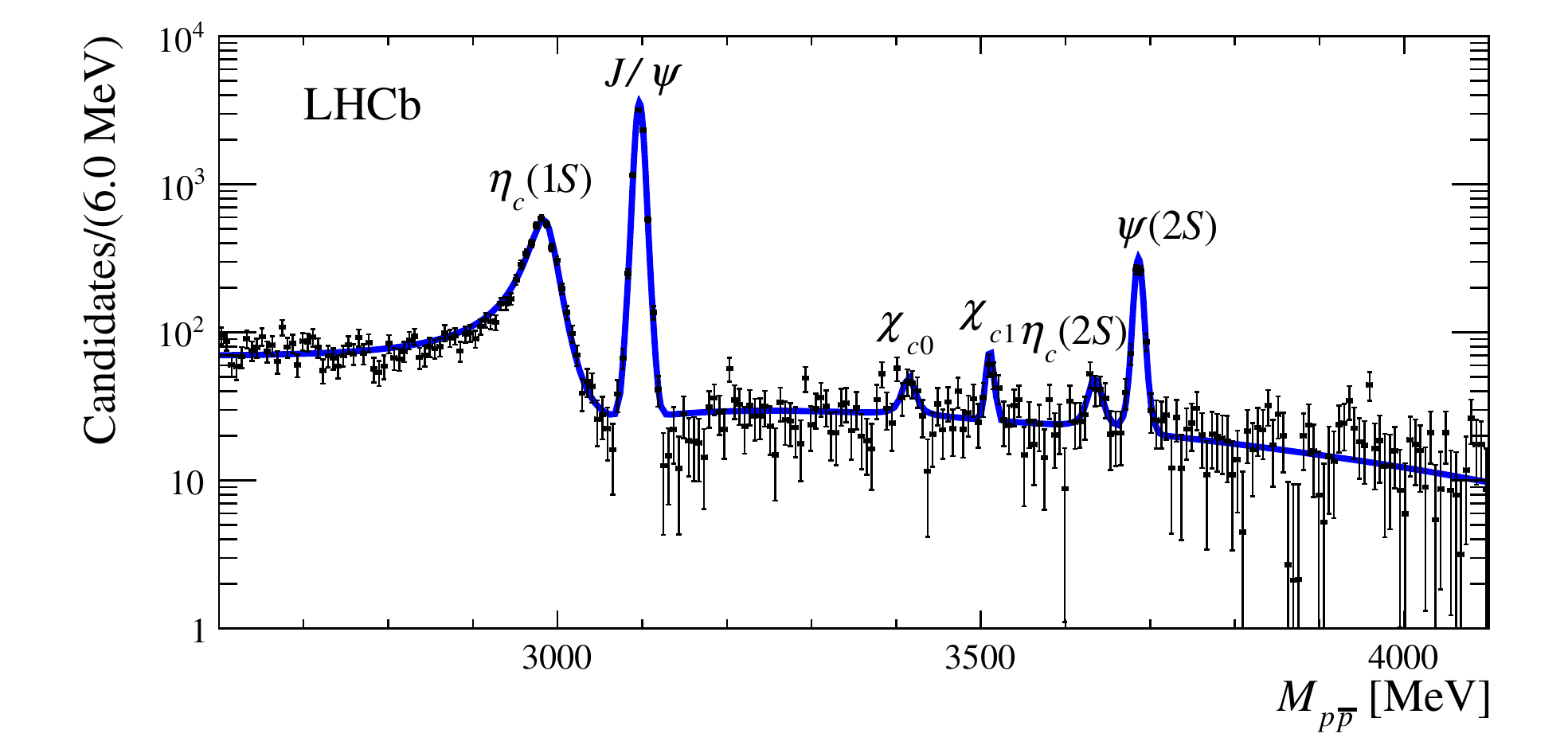}
\caption{Invariant mass spectrum of the $p \bar{p}$ candidates. Background in the $B^{+} \to p \bar p K^{+}$
  distribution is subtracted using the \sPlot technique as described
  in the text. The total fit
  curve is superimposed.}
\label{fig:pp}
\end{figure}

Six charmonium resonances are included in the nominal fit to the
$p\bar p$ invariant mass spectrum: $\eta_{c}(1S)$, $\jpsi$,
$\chi_{c0}$, $\chi_{c1}$, $\eta_{c}(2S)$ and $\psi(2S)$. Alternative fits including the
$\psi(3770)$ or the $X(3872)$ resonances are performed in order
to estimate upper limits on their branching fractions. The $\jpsi$
and $\psi(2S)$ peaks are parameterised with a
double Gaussian PDF. The $\eta_{c}(1S)$, $\eta_{c}(2S)$,
$\chi_{c0}$ and $\psi(3770)$ shapes are modelled with a
relativistic Breit-Wigner PDF convolved with a Gaussian PDF. The
$X(3872)$ and the $\chi_{c1}$  are described with a Gaussian
PDF. 
Due to the $B^{+}$ mass constraint in
the vertex fit, the $p \bar p$ mass resolution is effectively constant in the entire
$p\bar p$ spectrum. The
resolution for all charmonium states is fixed to that of the $\jpsi$ state. The masses of the $\chi_{c0}$,
$\chi_{c1}$, $X(3872)$ and $\psi(3770)$ states are fixed to the
known values~\cite{PDG2014}.
The $\jpsi$ and $\psi(2S)$ peak positions ($M_{\jpsi}$ and
$M_{\psi(2S)}$), the mass differences ($M_{\jpsi}-M_{\eta_{c}(1S)}$ and
$M_{\psi(2S)}-M_{\eta_{c}(2S)}$), and the natural width of the $\eta_c$(1S)
state ($\Gamma_{\eta_c({\rm 1S})}$) are free parameters and are obtained from the fit to the data. A Gaussian constraint to the
average value for the natural width of the $\eta_{c}(2S)$ is
applied~\cite{PDG2014}.
The $p \bar p$ non-resonant component is assumed to have no relative
orbital angular momentum, $J=0$.
The fit includes a possible interference effect between the
$\eta_{c}(1S)$ state and the $J=0$ non-resonant component. The 
amplitude is given by $|A|^{2} = |A_{\rm non \mbox{-}res} + f \, e^{i\delta} \,
A_{\eta_{c}(1S)}|^{2}$, where $A_{\rm non\mbox{-}res} $ is the amplitude of the
non-resonant component, $A_{\eta_{c}(1S)}$ is the amplitude of the
$\eta_{c}(1S)$ state, $\delta$ is the phase difference and $f$ a
normalisation factor. The shape of the non-resonant component in the
$p \bar p$ mass spectrum follows a phase-space distribution~\cite{PDG2014}. The fit result is shown in Fig.~\ref{fig:pp}.
Using Wilks' theorem~\cite{Wilks:1938dza}, the statistical significance for the $\eta_{c}(2S)$ signal is
found to be $6.4$ standard deviations.
No evidence for the  $\psi(3770)$ and $X(3872)$ resonances
is found. The signal yields are reported in Table~\ref{tab:yields}. 
\begin{table}[htbp]
\caption{Signal yields from the fit to the  $p \bar p$ mass spectrum
  in $B^{+} \to p \bar p K^{+}$ decays. The fit fractions of the $\eta_c(1S)$ and the non-resonant component in
the $J=0$ amplitude are $25\%$ and $65\%$ respectively.
 The fit fractions do not include uncertainties due to the
ambiguities in the relative phase of the interfering amplitudes. Uncertainties are
  statistical only.}
\centering
\begin{tabular}{c r@{}@{}l}
State & \multicolumn{2}{c}{Signal Yield}\\
\hline
$\eta_{c}(1S)$+non res.& $11246$ & $\,\pm\,119$ \\
$\jpsi$ & $6721$ & $\,\pm\,93$ \\
$\chi_{c0}$ & $84$ & $\,\pm\,22$\\
$\chi_{c1}$ & $95$ & $\,\pm\,16$\\
$\eta_{c}(2S)$ & $106$ & $\,\pm\,22$\\
$\psi(2S)$ & $588$ & $\,\pm\,30$\\
$\psi(3770)$ & $-6$  & $\,\pm\, 9$\\
$X(3872)$ & $-14$ & $\,\pm\,8$\\
\end{tabular}

\label{tab:yields}
\end{table}

\section{Efficiencies and systematic uncertainties}
The branching fraction of the $B^{+} \to [c\bar c] (\to p \bar p)
K^{+}$ decay for a specific $[c\bar c]$ resonance relative to that of the $\jpsi$ is given by 
\begin{equation}
{\cal R}_{[c\bar c]} \equiv \frac{{\mathcal B}(B^{+} \to [c\bar c] K^{+}) \times {\mathcal B}([c\bar c]\to
p \bar p)}{{\mathcal B}(B^{+} \to \jpsi K^{+}) \times {\mathcal B}(\jpsi \to
p \bar p)}  = \frac{N([c\bar c])}{N(\jpsi)}\times
\frac{\epsilon_{\jpsi}}{\epsilon_{c\bar c}}\label{eq:main},
\end{equation}
where $N([c\bar c]) \equiv  N(B^{+} \to [c\bar c] (\to p \bar p) K^{+})$ and 
 $N(\jpsi)  \equiv  N(B^{+} \to \jpsi (\to p \bar p) K^{+})$ are the
 numbers of decays and
 $\epsilon_{\jpsi}/\epsilon_{c\bar c}$ is the total efficiency ratio. The total efficiency is the product of the detector geometrical acceptance,
the trigger efficiency, the reconstruction and selection efficiency,
the PID efficiency, and the BDT classifier efficiency. The ratio of
the efficiencies between the signal and the normalising
$\jpsi$
channels is determined using simulated samples. To account for any
discrepancy between data and simulation, the PID efficiencies of kaons and protons are calibrated from data samples of $D^{*+} \to D^{0}
(\to K^{-}\pi^{+})
\pi^{+}$ and $\Lz^0
\to p
\pi^{-}$ decays. For each simulated candidate, its PID value is
replaced by a value extracted
randomly from the corresponding PID curves determined from control samples.
The selection is then applied to the PID-corrected simulated sample to estimate the efficiency.

Systematic uncertainties originate from the determination of the signal yields, efficiencies, selection procedure and branching fractions.
Since the final state is common for all considered decays, most of
the systematic uncertainties cancel in the ratios. 
Imperfect knowledge of the invariant mass distributions for the signal
and background causes systematic uncertainties
in the signal yield determination,  the mass difference and width measurements.
The contribution from the fit model is studied by using alternative
shapes for the $B^{+}$ component, for the $[c\bar c]$ states and for the background.
For the $B^{+}$ signal shape, a Gaussian PDF with power-law tails
on both sides and the
sum of two Gaussian PDFs with power-law tails are used as alternatives to the
Apollonios PDF. The combinatorial background
component in the $p\bar p K^{+}$ invariant mass is parameterised using a linear PDF. The effect of
removing the peaking background due to misidentified $B^{+} \to p
\bar p \pi^{+}$ decays is investigated by checking the variation of
the ratio of the 
branching fractions by including or neglecting this component in the
fit. Incorrect modelling of the partially
reconstructed background can also introduce a systematic
uncertainty. This is estimated by removing the $p \bar p K^{+}$
invariant mass fit range below $5.20 \gev$  in order to
exclude its contribution. In the fit to the $p\bar p$ spectrum, for the $\jpsi$ signal, the Apollonios PDF is
used as an alternative to the sum of two Gaussian PDFs. The range of the $p\bar p$ invariant mass
spectrum is also varied. The systematic uncertainty due to the
variation of the fit range gives a negligible contribution to the
branching fraction measurement while it is the largest contribution to
the $M_{\jpsi}-M_{\eta_{c}(1S)}$ difference. The largest variation in the ratio of the
branching fractions due to the fit model is assigned as the
corresponding systematic uncertainty. 

Possible biases related
to the signal selection criteria are
investigated by varying the BDT requirement and by checking the effect on the
branching fraction ratio and on the efficiency ratio, after accounting for statistical fluctuations. The maximum variation in the
ratio of the yields or the maximum variation in the mass
difference and width measurements are considered as an estimate of the
corresponding source of systematic uncertainty. In addition, variations in the procedure used to determine the
PID efficiency and the uncertainty due to the finite size of the
simulated samples, lead to an uncertainty on the efficiency ratio in the
the branching fractions evaluation. The total systematic uncertainties on the relative
branching fraction measurements, determined by adding the
individual contributions in quadrature, are listed in
Table~\ref{tab:totsyst}. 

\begin{table}[t!!!]
\caption{Systematic uncertainties in units of $10^{-4}$ on the $\eta_{c}(2S)$, $X(3872)$ and
  $\psi(3770)$ branching fraction measurements relative to that of the \jpsi. The efficiency
  contribution includes both the PID efficiency variation and the
  statistical error due to the finite size of the simulated samples.}
\centering
\begin{tabular}{cccc}
 & \multicolumn{1}{c}{$\eta_{c}(2S)$} & \multicolumn{1}{c}{$X(3872)$}  & \multicolumn{1}{c}{$\psi(3770)$}\\
\hline
Fit & 5 & 3 & 5  \\
BDT & 8  & 2 & 11 \\
Efficiency & 2 & 1  & 1 \\
\hline 
Total & 9 & 4 & 12\\
\end{tabular}
\label{tab:totsyst}
\end{table}

The significance, including systematic uncertainties, of the signals
is determined by convolving the profile likelihoods used in the yield 
determinations with a Gaussian with a width equal to the size of the 
systematic uncertainties that affect the yield. From the modified 
profile likelihood the significance of the $\eta_c(2S)$ signal is found to 
be $6.0$ standard deviations. The
upper limits at 90\% and 95\% confidence level on the $X(3872)$ and $\psi(3770)$ ratio of branching
fractions are determined from integrating the profile likelihood functions.

The measurements of the mass differences $M_{\jpsi}-M_{\eta_{c}(1S)}$ and
$M_{\psi(2S)}-M_{\eta_{c}(2S)}$ and the natural width of the
$\eta_{c}(1S)$ state are further affected by the uncertainty in the
momentum scale calibration. This systematic uncertainty is small for
the mass differences and negligible $(< 0.003 \mev)$ for the natural width.
Table~\ref{tab:systmassa} summarises the systematic uncertainties on the measurement of the  \mbox{$M_{\jpsi} -M_{\eta_{c}(1S)}$}, \mbox{$M_{\psi(2S)} -M_{\eta_{c}(2S)}$} mass differences and on the $\eta_{c}(1S)$ natural width. 
\begin{table}[t!!!]
\caption{Systematic uncertainties on the mass differences $M_{\jpsi}
  - M_{\eta_c(1S)}$, $M_{\psi(2S)} - M_{\eta_c(2S)}$ and the
  $\Gamma_{\eta_c(1S)}$ measurements. The systematic uncertainty associated to the momentum scale
  calibration is negligible for the total width $\Gamma_{\eta_c(1S)}$ measurement.}
\centering
\begin{tabular}{cccc}
 & $M_{\jpsi}-M_{\etac(1S)}$  &
 $M_{\psi(2S)} - M_{\eta_{c}(2S)}$
 & $\Gamma_{\eta_{c}(1S)}$ \\
& [\mev]   &
  [\mev] & [\mev]\\
\hline
Fit & 0.90 & 0.10  & 1.20 \\
BDT & 0.21  & 0.55  & 0.40  \\
Momentum scale & 0.03  & 0.06  &  - \\
\hline 
Total & 0.92 & 0.56  & 1.27 \\
\end{tabular}
\label{tab:systmassa}
\end{table}

\section{Results and conclusions}
A search for the $\eta_{c}(2S)$, $\psi(3770)$ and $X(3872)$ contributions in $B^{+} \to p \bar p K^{+}$ decays is performed using data corresponding to an integrated luminosity of $3.0\rm \, fb^{-1}$ recorded at centre-of-mass energies of $\sqrt{s}=7 \rm \, TeV$ and $8 \rm \, TeV$.
The branching fractions are determined using the $B^{+} \to \jpsi (\to p \bar p)K^{+}$ decay as normalization channel. 
The $\eta_{c}(2S) \to p \bar p$ decay is observed for the first time
with a total significance of $6.0$ standard deviations. The relative branching fraction is measured to be
\begin{equation*}
\frac{{\mathcal B}(B^{+} \to \eta_{c}(2S) K^{+}) \times {\mathcal B}(\eta_{c}(2S) \to
p \bar p)}{{\mathcal B}(B^{+} \to \jpsi K^{+}) \times {\mathcal B}(\jpsi \to
p \bar p)}  =\\ (1.58 \pm 0.33 \pm 0.09)\times 10^{-2},
\end{equation*}
where the first uncertainty is statistical and the second systematic.
For the $B^{+} \to X(3872) (\to p \bar p)K^{+}$ and the $B^{+} \to
\psi(3770) (\to p \bar p)K^{+}$ decays, the upper limits at 90 (95)\% confidence level  are
\begin{equation*}
\frac{{\mathcal B}(B^{+} \to \psi(3770) K^{+}) \times {\mathcal B}(\psi(3770)\to p \bar p)}{{\mathcal B}(B^{+} \to J/\psi K^{+}) \times {\mathcal B}(\jpsi \to
p \bar p)}  <\\
9 \, (10) \times 10^{-2},
\end{equation*}
\begin{equation*}
\frac{{\mathcal B}(B^{+} \to X(3872) K^{+}) \times {\mathcal B}(X(3872)\to p \bar p)}{{\mathcal B}(B^{+} \to \jpsi K^{+}) \times {\mathcal B}(\jpsi \to
p \bar p)}  <\\
0.20 \, (0.25) \times 10^{-2}.
\end{equation*}

The visible branching fraction calculated using the value of
${\mathcal B}(B^{+} \to \jpsi K^{+}) \times {\mathcal B}(\jpsi \to p
\bar p) = (2.2 \pm 0.1) \times 10^{-6}$~\cite{PDG2014} is determined to be
\begin{equation*}
{\mathcal B}(B^{+} \to \eta_{c}(2S) K^{+}) \times {\mathcal B}(\eta_{c}(2S) \to
p \bar p)  =\\ (3.47 \pm 0.72 \pm 0.20
\pm 0.16) \times 10^{-8},
\end{equation*}
where the last uncertainty is due to the uncertainty on ${\mathcal B}(B^{+} \to
\jpsi K^{+}) \times {\mathcal B}(\jpsi \to p \bar p)$.

The differences between $M_{\jpsi}$ and $M_{\eta_{c}(1S)}$ and between $M_{\psi(2S)}$ and $M_{\eta_{c}(2S)}$ are measured
to be
\begin{equation*}
M_{\jpsi} - M_{\eta_{c}(1S)} = 110.2 \pm 0.5 \pm
0.9  \mev,
\end{equation*}
\begin{equation*}
M_{\psi(2S)} - M_{\eta_{c}(2S)}= 52.5 \pm 1.7 \pm 0.6\mev.
\end{equation*}
The natural width of the $\eta_{c}(1S)$ is found to be
\begin{equation*}
\Gamma_{\eta_{c}(1S)} = 34.0 \pm 1.9  \pm 1.3 \mev.
\end{equation*}
In contrast to the determinations using radiative decays, these mass and
width determinations do not depend on the knowledge of the line shapes
of the magnetic dipole transition.

\section*{Acknowledgements}
\noindent We express our gratitude to our colleagues in the CERN
accelerator departments for the excellent performance of the LHC. We
thank the technical and administrative staff at the LHCb
institutes. We acknowledge support from CERN and from the national
agencies: CAPES, CNPq, FAPERJ and FINEP (Brazil); NSFC (China);
CNRS/IN2P3 (France); BMBF, DFG and MPG (Germany); INFN (Italy); 
FOM and NWO (The Netherlands); MNiSW and NCN (Poland); MEN/IFA (Romania); 
MinES and FANO (Russia); MinECo (Spain); SNSF and SER (Switzerland); 
NASU (Ukraine); STFC (United Kingdom); NSF (USA).
We acknowledge the computing resources that are provided by CERN, IN2P3 (France), KIT and DESY (Germany), INFN (Italy), SURF (The Netherlands), PIC (Spain), GridPP (United Kingdom), RRCKI and Yandex LLC (Russia), CSCS (Switzerland), IFIN-HH (Romania), CBPF (Brazil), PL-GRID (Poland) and OSC (USA). We are indebted to the communities behind the multiple open 
source software packages on which we depend.
Individual groups or members have received support from AvH Foundation (Germany),
EPLANET, Marie Sk\l{}odowska-Curie Actions and ERC (European Union), 
Conseil G\'{e}n\'{e}ral de Haute-Savoie, Labex ENIGMASS and OCEVU, 
R\'{e}gion Auvergne (France), RFBR and Yandex LLC (Russia), GVA, XuntaGal and GENCAT (Spain), Herchel Smith Fund, The Royal Society, Royal Commission for the Exhibition of 1851 and the Leverhulme Trust (United Kingdom).

\addcontentsline{toc}{section}{References}
\setboolean{inbibliography}{true}
\bibliographystyle{LHCb}
\bibliography{main,LHCb-PAPER,LHCb-CONF,LHCb-DP,LHCb-TDR}

\newpage
 
\newpage
\centerline{\large\bf LHCb collaboration}
\begin{flushleft}
\small
R.~Aaij$^{39}$,
B.~Adeva$^{38}$,
M.~Adinolfi$^{47}$,
Z.~Ajaltouni$^{5}$,
S.~Akar$^{6}$,
J.~Albrecht$^{10}$,
F.~Alessio$^{39}$,
M.~Alexander$^{52}$,
S.~Ali$^{42}$,
G.~Alkhazov$^{31}$
% ...

{\footnotesize \it
% ...
$ ^{5}$Clermont Universit{\'e}, Universit{\'e} Blaise Pascal, CNRS/IN2P3, LPC, Clermont-Ferrand, France\\
$ ^{6}$CPPM, Aix-Marseille Universit{\'e}, CNRS/IN2P3, Marseille, France\\
% ...
$ ^{10}$Fakult{\"a}t Physik, Technische Universit{\"a}t Dortmund, Dortmund, Germany\\
% ...
$ ^{31}$Petersburg Nuclear Physics Institute (PNPI), Gatchina, Russia\\
% ...
$ ^{38}$Universidad de Santiago de Compostela, Santiago de Compostela, Spain\\
$ ^{39}$European Organization for Nuclear Research (CERN), Geneva, Switzerland\\
% ...
$ ^{42}$Nikhef National Institute for Subatomic Physics, Amsterdam, The Netherlands\\
% ...
$ ^{47}$H.H. Wills Physics Laboratory, University of Bristol, Bristol, United Kingdom\\
% ...
$ ^{52}$School of Physics and Astronomy, University of Glasgow, Glasgow, United Kingdom\\
% ...
}
\end{flushleft}

\end{document}